\pdfoutput=1

\documentclass[conference]{IEEEtran}
\ifCLASSINFOpdf
\else
\fi
\usepackage{tkz-graph}
\usepackage{pgf-pie}
\usepackage{geometry} 
\usepackage{amsmath}
\usepackage{cleveref}
\usepackage{tikz}
\usepackage{graphicx}
\usepackage[rightcaption]{sidecap}
\usepackage{dblfloatfix}
\usepackage{float}
\usepackage{array}

\usepackage[normalem]{ulem}

\usetikzlibrary{positioning, shapes.geometric}

\definecolor{dark green}{HTML}{FFA500}
\definecolor{dark red}{HTML}{800080}
\definecolor{gray}{HTML}{c4c4c4}
     
\begin{document}
%
\title{Optical Routing with Binary Optimisation and Quantum Annealing }
\setlength{\belowcaptionskip}{-30pt}


\author{
\IEEEauthorblockN{Ethan Davies\IEEEauthorrefmark{1},
Darren Banfield\IEEEauthorrefmark{2},
Vlad C\u{a}rare\IEEEauthorrefmark{3},
Ben Weaver\IEEEauthorrefmark{4},
Catherine White\IEEEauthorrefmark{5} and
Nigel Walker\IEEEauthorrefmark{5}
}
\IEEEauthorblockA{\IEEEauthorrefmark{1}Information Security Group, Royal Holloway, Univ. of London\\
Email: Ethan.Davies.2021@live.rhul.ac.uk
\\
}
\IEEEauthorblockA{\IEEEauthorrefmark{2}Dep't Mathematics, Royal Holloway, Univ. of London\\
Email: Darren.Banfield.2020@live.rhul.ac.uk\\}
\IEEEauthorblockA{\IEEEauthorrefmark{3}Dep't Physics,
University of Cambridge\\
}
\IEEEauthorblockA{\IEEEauthorrefmark{4}National Quantum Computing Centre, Rutherford Appleton Laboratory, Didcot\\
}
\IEEEauthorblockA{\IEEEauthorrefmark{5}BT Research,
Adastral Park, Ipswich\\
}
}




\maketitle

\begin{abstract}
A challenge for scalability of demand-responsive, elastic optical Dense Wavelength Division Multiplexing (DWDM) and Flexgrid networks is the computational complexity of allocating many optical routes on large networks. We demonstrate that demand satisfaction problems in communication networks can be formulated as quadratic unconstrained binary optimisation (QUBO) problems, and solved using a hybrid quantum annealer. Efficient encodings are developed which solve both unicast and multicast multicommodity-flow problems, while also adhering to individual requirements for maximum latency and resilience for each route. We present several QUBO formulations and analyse the qubit scaling. We demonstrate solutions using a hybrid solver, D-Wave Quantum Advantage QPU. Progress in generating optimal solutions 
with efficient use of computational resources will be beneficial to telecoms operators, enabling them to run dynamic optical network infrastructures which use resources efficiently, are resilient to local faults and cyber-attacks, and can be elastically responsive to demands.
\end{abstract}


\section{Introduction}
Techniques for generating sets of routes in communication networks have many applications. For example, elastic optical networking and dynamic power utilisation require fast solution, at low computational cost. 

Models of networks at different network layers have different characteristics. Optical transmission uses a discrete frequency grid, so presents a discrete integer optimisation problem. Algorithmically, optimising multi-commodity discrete flows through a graph is NP Hard \cite{even_complexity_1976}, which can in many instances be intractable - particularly for large networks over which many routes must be planned, and where the capacity use is close to the `crunch' point.

New paradigms, such as Flexgrid and optical multicast \cite{simmons_routing_2020}, expand the domain of optical routing problems. We have developed a framework to encode all such problems as quadratic unconstrained binary optimisation (QUBO) problems. This form is compatible with new computational platforms including quantum annealers, hybrid solvers \cite{McGeoch_Pau_William_2020}, boson samplers \cite{bradler_binary_2023} and Variational Quantum Eigensolver \cite{Glos_Krawiec_Zimborás_2022}. There is existing research applying QUBO to other types of real world network optimisation, such as electrical power distribution \cite{silva_quantum_2023}.

\section{The Problem}
\label{The Problem}

Real world telecommunication networks can be represented as a directed graphs, where graph nodes are the locations in which the information may be generated, routed and consumed and directed edges are the communication links, bounded by capacity. Connectivity between nodes is usually incomplete and so information must be routed through intermediate nodes. Routes should satisfy demands (such as linking sources and sinks with sufficient capacity), satisfy constraints such as maximum latency, and use the network efficiently. To ensure resilient communication between two nodes, we also need to identify disjoint routes, so that if a single resource is incapacitated (e.g. by rodents, construction, or cyber-attacks), one route remains. 

Let $G$ be a weighted directed graph representing the communication network, with vertices $V$ and edges $E$ with associated edge parameters such as the economic cost of using links, link-latencies and link-capacities. Let $K$ be a set of labels called commodities (representing the diverse information demands to be carried over the network). For each commodity $k$, a subset $S_k \subset V$ of vertices are the sources, and a subset $T_k \subset V$ of vertices are the sinks.

A valid route set is a set $\Gamma$ of routes in $G$ such that for every sink vertex $v \in T$, there exists a source $s \in S$ labelled with the same commodity, and a route $\gamma \in \Gamma$ such that $\gamma$ is a route from $s$ to $v$. The latency of the route $\gamma$ is the sum of the latencies of the edges in route $\gamma$, and the latency of the route set $\Gamma$ is the maximum of the latencies of the routes within $\Gamma$. Goals for the valid route set may include satisfying conditions associated with the weight dimensions (such as not exceeding link capacity) and  optimisations (such as providing lowest possible latency). We may wish to minimise the maximum of the latencies of the route sets $\Gamma_{1}$ and $\Gamma_{2}$, or simply require that this maximum is below a certain threshold. We may also wish for some financial or resource costs of a route set to be minimised or below some threshold.

In order to provide resilience, we seek valid routes 
which satisfy a disjointness condition. This condition could vary between problems, but in developing our algebraic framework, we consider the following:
\begin{enumerate}
\item{Edge disjoint routes. 
}
\item{Node disjoint routes. 
}
\item {Shared Risk Group (SRG) disjoint routes. More generally, define SRGs to be groups of edges and nodes. All routes in $\Gamma_{1}$ are SRG disjoint with all routes in $\Gamma_{2}$.}
\end{enumerate}
\emph{Important}: Finding resilient optical routes can be divided into two steps: firstly finding routes that meet non-spectral demands, resilience requirements and constraints (an NP hard problem which we solve in Sections \ref{subsec:time_oriented} and \ref{sec:sink_oriented}); and secondly, assigning non-blocking frequencies to the routes. However if the network has spectral congestion, it is necessary to solve the routing and spectral assignment problem simultaneously (see Section \ref{sec:wavelength_assignment}).

\subsection{Toy Problems} 
\begin{figure}[t]
\begin{center}


\scalebox{0.57}{
\begin{tikzpicture}[>=latex]
\SetGraphUnit{2}
\tikzset{VertexStyle/.style = {shape = circle,
draw = black,
inner sep = 2pt,%
minimum size = 5mm,
outer sep = 0pt,
fill = green!60}}
\Vertex[Math, L=S_1]{S}
\tikzset{VertexStyle/.style = {shape = circle,
draw = black,
inner sep = 2pt,%
minimum size = 5mm,
outer sep = 0pt,
fill = white!60}}
\EA[NoLabel](S){A}
\EA[NoLabel](A){B}
\NO[NoLabel](S){D}
\EA[NoLabel](D){F}
\EA[NoLabel](F){G}
\EA[NoLabel](G){H}
\NO[NoLabel](S){D}
\EA[NoLabel](D){F}
\EA[NoLabel](F){G}
\EA[NoLabel](G){H}
\NO[NoLabel](D){I}
\EA[NoLabel](I){J}
\EA[NoLabel](J){K}
\EA[NoLabel](K){L}
\NO[NoLabel](J){M}
\EA[NoLabel](M){N}

\tikzset{VertexStyle/.style = {shape = circle,
draw = black,
inner sep = 2pt,%
minimum size = 5mm,
outer sep = 0pt,
fill = red!60}}
\EA[Math, L=T_1](B){C}
\NO[Math, L=T_1](D){I}
\NO[Math, L=T_1](J){M}
\EA[Math, L=T_1](N){O}
\tikzset{EdgeStyle/.style = {->, line width= 2.5pt, color = gray}}
 \Edge[label=$1$](A)(D)
 \Edge[ label=$2$](K)(F)
 \Edge[label=$1$](L)(N)

\tikzset{EdgeStyle/.style = {->, line width= 2.5pt, color = blue}}
\Edge[label=$5$](S)(D)
\Edge[label=$2$](D)(I)
\Edge[label=$2$](D)(F)
\Edge[label=$3$](F)(B)
\Edge[label=$4$](B)(C)
\Edge[label=$2$](B)(H)
\Edge[label=$2$](H)(K)
\Edge[label=$3$](K)(M)
\Edge[label=$1$](N)(O)
\Edge[label=$5$](K)(N)

\tikzset{EdgeStyle/.style = {->, line width= 2.5pt, color = red}}
\Edge[style = {bend right},label=$1$](G)(J)
\Edge[label=$2$](J)(I)
\Edge[label=$1$](J)(M)
\Edge[label=$2$](H)(L)
\Edge[label=$5$](L)(O)
\Edge[label=$2$](S)(A)
\Edge[label=$4$](A)(F)
\Edge[label=$3$](F)(G)
\Edge[label=$1$](G)(H)
\Edge[label=$2$](H)(C)

\tikzset{VertexStyle/.style = {shape = circle,
draw = black,
inner sep = 2pt,%
minimum size = 5mm,
outer sep = 0pt,
fill = green!60}}
\tikzset{VertexStyle/.style = {shape = circle,
draw = white,
inner sep = 2pt,%
minimum size = 5mm,
outer sep = 0pt,
fill = white!60}}

\SOEA[NoLabel](L){temp1}
\SO[NoLabel](temp1){temp2}

\tikzset{VertexStyle/.style = {shape = circle,
draw = black,
inner sep = 2pt,%
minimum size = 5mm,
outer sep = 0pt,
fill = red!60}}

\EA[Math, L=T_1](temp2){A1}
\tikzset{VertexStyle/.style = {shape = circle,
draw = black,
inner sep = 2pt,%
minimum size = 5mm,
outer sep = 0pt,
fill = red!60}}
\EA[Math, L=T_2](A1){B1}

\tikzset{VertexStyle/.style = {shape = circle,
draw = black,
inner sep = 2pt,%
minimum size = 5mm,
outer sep = 0pt,
fill = white!60}}
\NOWE[NoLabel](A1){C1}
\EA[NoLabel](C1){D1}
\EA[NoLabel](D1){E1}
\EA[NoLabel](E1){F1}
\NO[NoLabel](C1){G1}

\tikzset{VertexStyle/.style = {shape = circle,
draw = black,
inner sep = 2pt,%
minimum size = 5mm,
outer sep = 0pt,
fill = green!60}}

\NO[Math, L=S_1](C1){G1}
\tikzset{VertexStyle/.style = {shape = circle,
draw = black,
inner sep = 2pt,%
minimum size = 5mm,
outer sep = 0pt,
fill = white!60}}

\EA[NoLabel](G1){H1}
\tikzset{VertexStyle/.style = {shape = circle,
draw = black,
inner sep = 2pt,%
minimum size = 5mm,
outer sep = 0pt,
fill = green!60}}

\EA[Math, L=S_2](H1){I1}

\tikzset{VertexStyle/.style = {shape = circle,
draw = black,
inner sep = 2pt,%
minimum size = 5mm,
outer sep = 0pt,
fill = white!60}}

\EA[NoLabel](I1){J1}
\NO[NoLabel](G1){K1}
\EA[NoLabel](K1){L1}
\tikzset{VertexStyle/.style = {shape = circle,
draw = black,
inner sep = 2pt,%
minimum size = 5mm,
outer sep = 0pt,
fill = red!60}}
\EA[Math, L=T_1](L1){M1}

\tikzset{VertexStyle/.style = {shape = circle,
draw = black,
inner sep = 2pt,%
minimum size = 5mm,
outer sep = 0pt,
fill = white!60}}

\EA[NoLabel](M1){N1}
\tikzset{VertexStyle/.style = {shape = circle,
draw = black,
inner sep = 2pt,%
minimum size = 5mm,
outer sep = 0pt,
fill = red!60}}
\NO[Math, L=T_1](K1){O1}
\tikzset{VertexStyle/.style = {shape = circle,
draw = black,
inner sep = 2pt,%
minimum size = 5mm,
outer sep = 0pt,
fill = red!60}}
\EA[Math, L=T_2](O1){P1}

\tikzset{VertexStyle/.style = {shape = circle,
draw = black,
inner sep = 2pt,%
minimum size = 5mm,
outer sep = 0pt,
fill = white!60}}

\NO[NoLabel](M1){Q1}
\tikzset{VertexStyle/.style = {shape = circle,
draw = black,
inner sep = 2pt,%
minimum size = 5mm,
outer sep = 0pt,
fill = green!60}}

\tikzset{EdgeStyle/.style = {->, line width= 2.5pt, color = blue}}
\Edge[label = $1$](C1)(A1)
\Edge[label = $5$](G1)(C1)

\Edge[label = $1$](L1)(M1)
\Edge[label = $4$](L1)(O1)

\tikzset{EdgeStyle/.style = {->, dashed,line width= 2.5pt, color = blue}}
\Edge[label = $1$](I1)(F1)
\Edge[label = $4$](F1)(B1)
\Edge[label = $2$](I1)(L1)
\Edge[label = $5$](L1)(P1)

\tikzset{EdgeStyle/.style = {->, line width= 2.5pt, color = red}}
\Edge[label = $5$](G1)(K1)
\Edge[label = $3$](K1)(O1)
\Edge[label = $4$](G1)(D1)
\Edge[label = $3$](D1)(A1)
\Edge[label = $2$](D1)(E1)
\Edge[style = {bend right}, label = $2$](E1)(J1)
\Edge[label = $2$](J1)(M1)

\tikzset{EdgeStyle/.style = {->,dashed, line width= 2.5pt, color = red}}
\Edge[label = $3$](I1)(E1)
\Edge[label = $2$](E1)(B1)
\Edge[label = $1$](J1)(N1)
\Edge[label = $2$](N1)(Q1)
\Edge[label = $2$](Q1)(P1)

\tikzset{EdgeStyle/.style = {->, line width= 2.5pt, color = gray}}
\Edge[label = $3$](Q1)(L1)
\Edge[label = $4$](N1)(M1)
\Edge[label = $1$](F1)(E1)
\Edge[label = $4$](H1)(D1)
\Edge[label = $3$](C1)(D1)
\Edge[label = $1$](K1)(H1)
\Edge[label = $3$](H1)(E1)

\tikzset{EdgeStyle/.style = {->, line width= 2.5pt, color = blue}}
\Edge[style = {bend right}, label = $5$](G1)(L1)
\end{tikzpicture}}

\caption{
In this diagram, S are sources, T are sinks, Labels on edges indicate latency of the link (arb. units).
\textbf{Problem A} (left) - Single multicast, the goal is to find two edge-disjoint multicast trees, with latency $\leq$ 20. 
\textbf{Problem B} (right) - Two simultaneous multicasts on non-planar graph, the goal is to find, for each, a pair of node-resilient multicast trees that connects each source to its respective sinks with latency $\leq$ 10.}
\label{Toy problems solution}
\end{center}
\vspace{-5mm}
\end{figure}
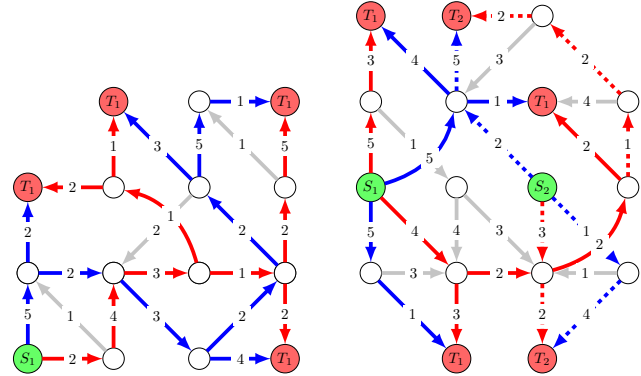

We investigate the optical multicast routing problem. (Unicast can be considered a special case, with one sink).
Figure \ref{Toy problems solution} shows two model graph topologies and routing problems. \emph{Problem A} has a single optical multicast source (green), and four sinks (red). The problem is to identify two edge-disjoint resilient multicast routes from the source to all four sinks. The two routes (blue and red) have no edges in common, and each pass from the source to both sinks. In addition, the maximum latency of the two multicast routes is $\leq$ 20. \emph{Problem B} involves planning more than one optical multicast, (multicommodity routing). There are two different multicast sources (green) each with a different set of sinks (red). For each, the goal is to identify two (resilient) node-disjoint, multicast routes from each source to its sinks, and additionally such that the maximum latency to any sink is $\leq$ 10.
\vspace{-1mm}

\section{Quantum Annealing Approach}

Quantum annealing is a computational paradigm in which the energy of a coupled system is used to represent a parameter to be minimised in a solution \cite{Adiabatic_Albash_2018}. The physical dynamical evolution of qubits (quantum representations of information bits) is governed by a Hamiltonian (energy dynamics equation). The qubits are prepared in a ground state of an initial Hamiltonian $H_{0}$ which is evolved under an applied physical field, to a desired Hamiltonian $Q$ with local field values representing the optimisation problem, for which the minimum (ground) state is required. Provided this evolution is performed adiabatically (in practice, sufficiently slowly), then the resulting state should have a high probability of being in, or close to, the ground state of $Q$. Satisfiability problems can also be formulated by encoding them in as an optimisation that is minimised by constraint satisfaction. 


We therefore need a definition of $Q$ for which the minimum energy solution provides a solution to optical routing problems. The current generation of quantum annealers can solve problems in QUBO form. QUBO form involves minimising the value of $\mathbf{x}^{\dagger} Q \mathbf{x}$
where $\mathbf{x}$ is a binary vector and $Q$ is a real symmetric matrix. Each coordinate $x_{j}$ in the vector corresponds to the state of the single qubit $j$.

Despite the requirements for a binary input vector $\mathbf{x}$, and the restriction to quadratic terms, a many problems can be written as QUBOs. Key to this approach is writing constraints as quadratic terms; we describe some of the tools in section \ref{sec:QUBObuilding}.

\emph{Important}: The analysis of qubit scaling in this work applies to pure quantum annealing. Due to current restrictions on QPU size and performance, we have however used the hybrid solver for the experimental section. This uses a combination of classical heuristics and quantum annealing, and therefore ``qubit scaling" is here ``binary variable scaling". 

\small
\begin{table*}[!b]
  \centering
\textsc{TABLE II}\\
\textsc{Summary of constraints represented in the time-oriented approach.}
\\ \vspace{0.3cm}
\begin{tabular}{ |m{5cm}|m{9.9cm}|}
 \hline
 \hfil Equation & \hfil Interpretation\\
 \hline
\hfil  $\forall a \in S_k\cup T_k\quad x_{i,k,a} = 1$& Sources and sinks must appear in the trees for their commodities. Since we know the values of these variables they can be replaced as constants in subsequent constraints, thus reducing the number of variables needed.\\
\hline 
\hfil $\sum_{a} e_{i,k,(a,b)} = x_{i,k,b}$& Node $b$ is connected to a tree if there is exactly one edge leading to $b$, otherwise the node is not in the tree and no edges lead to $b$.\\
\hline
\hfil $e_{i,k,(a,b)}\leq x_{i,k,a}$& A directed edge can only be used if the source node is in the tree.\\
\hline
\hfil $\xi_{i,k,(a,b),j}= e_{i,k,(a,b)}  t_{i,k,b,j} $ & $\xi_{i,k,(a,b),j}$ is the product of the two binary terms. These variables are used to construct the integer $e_{i,k(a,b)}t_{i,k,b}= \sum_i 2^i\xi_{i,k,(a,b),j}$.\\
\hline
\hfil $t_{i,k,a} \geq e_{i,k,(a,b)}T_{(a,b)} +e_{i,k,(a,b)}t_{i,k,b} $& If edge $(a,b)$ is used, the time for information to flow from $a$ to any node $c$ beyond $b$  must be at least the time taken to flow from $b$ to $c$ plus the time taken to travel from $a$ to $b$. (This prevents loops and ensures the route is a tree). This constraint requires an additional $2 K E \log(T_{max})$ slack variables.\\
\hline
\hfil $\forall a \in S_k\quad t_{i,k,a}\leq T_{max}$& The time taken for the source to reach all sinks is constrained to be at most a maximum total latency $T_{max}$. This would require an additional $2K\langle S\rangle\log(T_{max})$ variables. Another approach we use for this constraint, is to instead encode it directly by adjusting the binary expansion of each time variable so that it may not exceed $T_{max}$.\\
\hline
\hfil $\sum_k c_k e_{i,k,(a,b)}\leq C_{(a,b)}$ & The total capacity used along an edge is less than the total capacity of the edge. This uses an extra $2E\langle \log(C)\rangle$ slack variables\\
\hline
\hfil $ \forall a, (a,b) \in SRG_j\quad x_{i,k,a}\leq SRG_{i,j}$ \newline
\textcolor{white}{easteregg!!!!!!!!!!!!!}
$e_{i,k,(a,b)} \leq SRG_{i,j}$ & If a node or edge in $SRG_j$ is used in solution $i$, then $SRG_{i,j}=1$\\
\hline
\hfil $SRG_{0,j}+SRG_{1,j}\leq 1$ & Both solutions cannot use elements in the same risk group $SRG_j$.\\
\hline 
\hfil $e_{i,k,(a,b)}\leq e_{i,(a,b)}$ & If a commodity uses an edge $(a,b)$ for solution $i$, then $e_{i,(a,b)}=1$\\
\hline
\hfil $\sum_{(a,b)\in E} w_{(a,b)}e_{i,(a,b)} \leq W$ & The total cost of route set $i$ is below the cost threshold.\\
\hline
\end{tabular}
\label{table:constraints}
\end{table*}
\normalsize

\section{QUBO Construction}
\label{sec:QUBObuilding}

The approach to constructing a QUBO instance of a desired problem is to break the problem down into smaller constraints, which we wish to be satisfied simultaneously. 
Suppose $x^*$ minimises the $x^{*\dagger}Q_ix^*$ for $i \in \{0,...,n\}$. Then $\forall \, \lambda_i \geq 0, \,x^*$ also minimises $x^{*\dagger} \left(\sum_i \lambda_i Q_i \right) x^*$. 

The values $\lambda_{i}$ are hyperparameters which can be freely tuned (independent of the problem constraints) to weight some constraints more highly than others. 
The most important constraints should be given higher weights. However, none of the weights should be too small relative to the others; if so, the noise in the quantum system may result in this constraint not being satisfied. This tuning stage can be fixed based on prior experiment or theory, or adjusted iteratively during experiments to improve the quality of results.

Our problems use certain constraint categories which can be transformed into the QUBO formalism as follows:

\begin{itemize}
    \item For two binary variables $x,y$, we can constrain them such that $x+y\leq 1$ with the quadratic expressions $$Q=xy.$$
    \item For two binary variables $x,y$, we can constrain them such that $x\leq y$, with the quadratic expression $$Q=x - xy.$$
    \item To access higher power terms than quadratic, it is possible to introduce extra variables $z:= x\cdot y$ and constrain $z$ to be the product of $x$ and $y$. This can be achieved with the quadratic expression $$Q = xy-2xz-2yz+3z.$$
    \item To obtain bounded integer variables, we can combine binary variables. We write $x= \sum_i 2^i x_i$ where each $x_i \in  \{0,1\} $. If we instead want to fix $x$ to be below some quantity $C$, we can adjust the final term in the binary expansion, such that $x$ cannot exceed $C$.
    \item Constraints on linear combinations of variables can be achieved. The constraint $\sum_i \alpha_i x_i=C$ can be represented with the quadratic expression $$Q=(\sum_i \alpha_i x_i - C)^2.$$
    \item Inequalities can also be encoded in a similar fashion. The requirement $\sum_i \alpha_i x_i \leq C$ is equivalent to the existence of a non-negative $\xi$ such that $\sum_i \alpha_i x_i+\xi=C$. We create extra binary variables for the expansion of $\xi$ and obtain 
    $$Q=(\sum_i \alpha_i x_i + \sum_i 2^i \xi_i-C)^2.$$ This requires the introduction of additional $\mathcal{O}\left(\log(C-\sum_{i:\alpha_i<0}\alpha_i)\right)$ slack variables, for cases where $\alpha_i<0$.
\end{itemize}

\section{QUBO Formulation} \label{QuboFormalism}

In this section we use the following notation for the problem defined in \cref{The Problem}:
\begin{itemize}
    \item A graph $G=(V,E)$ where each edge $(a,b)$ has an associated latency, $T_{(a,b)}$, capacity, $C_{(a,b)}$ and cost, $w_{(a,b)}$.
    \item A set of commodities $K$, with associated demand $c_k$ and sources $S_k$ and sinks $T_k$.
    \item A maximum time $T_{max}$ by which time all information must be transmitted.
    \item $SRG_j$, The Shared Risk Groups, indexed by j.  
\end{itemize}
The goal is to obtain a pair of resilient solutions such that all commodities can reach the required sinks within the required time while the capacity of any edge is not exceeded. 
Our approach is to construct a multicast tree (or forest for multiple sources) such that the time taken for any route is below $T_{max}$. We add further constraints to ensure the capacity of any edge is not violated. From this point, we ensure the required resilience properties by generating two independent solutions. To do this, we take two (or more) instances of the variables representing the network configuration, and create further constraints between these instances which enforce finding independent solutions to the problem. Depending on the nature of the resilience constraints, this may enforce edge, node or SRG-disjoint resilience.

\begin{SCfigure}
\scalebox{0.6}{
\begin{tikzpicture}[>=latex]
\SetGraphUnit{2.5}
\tikzset{VertexStyle/.style = {shape = circle,
draw = black,
inner sep = 2pt,%
minimum size = 5mm,
outer sep = 0pt,
fill = green!60}}
\Vertex{S}
\tikzset{VertexStyle/.style = {shape = circle,
draw = black,
inner sep = 2pt,%
minimum size = 5mm,
outer sep = 0pt,
fill = cyan!60}}
\EA(S){A}
\NO(S){C}
\EA(A){B}
\EA(C){D}

\tikzset{VertexStyle/.style = {shape = circle,
draw = black,
inner sep = 2pt,%
minimum size = 5mm,
outer sep = 0pt,
fill = red!60}}
\NO(C){E}

\tikzset{EdgeStyle/.style = {->, dashed, line width= 2.5pt, color = gray}}
\Edge[label=$5$](S)(A)
\Edge[ label=$4$](S)(C)
\Edge[label=$1$](B)(D)
\Edge[label=$1$](C)(E)
 \Edge[label=$1$](A)(B)
 
\tikzset{EdgeStyle/.style = {->, line width= 2.5pt, color = blue}}
\Edge[label=$1$](C)(D)
\Edge[label=$1$](D)(A)
\Edge[label=$1$](A)(C)
\Edge[label=$1$](D)(E)

\end{tikzpicture}}
\qquad

\caption{Illustration of an invalid solution generated due to the existence of loops in the network. This is not a valid route since there is no connection from the source. The constraints in row 5 of Table \textsc{II} and row 3 of Table \textsc{IV} prevent these loops.}
\label{fig:loops}
\end{SCfigure}
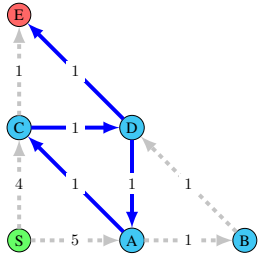

\small
\begin{table}
\footnotesize 
\centering
\textsc{TABLE I}\\
\textsc{Binary variables used in the time-oriented QUBO described in \cref{subsec:time_oriented}.}
\\ \vspace{0.3cm}
\begin{tabular}{|m{1.2cm}|m{3.4cm}|m{2.3cm}|}
 \hline
 \hfil Variable & \hfil Interpretation & \hfil \# of Variables\\
 \hline
 \hfil $x_{i,k,a}$   &  Node $a$ is used in the tree for commodity $k$ for solution $i$.    & \hfil $2 K V$\\
 \hline
 \hfil $e_{i,k,(a,b)}$&   Edge $(a,b)$ is used in the tree for commodity $k$ for solution $i$.  & \hfil $2 K E$\\
 \hline
 \hfil $e_{i,(a,b)}$&   Edge $(a,b)$ is used in solution $i$.  & \hfil $2E$\\
 \hline
 
 \hfil $t_{i,k,a,j}$ &  Time variables for each node, $$\sum_j 2^j t_{i,k,a,j} = t_{i,k,a}.$$ & \hfil $2  K V \log(T_{max})$\\
 \hline
  $\xi_{i,k,(a,b),j}$ &  The product of $e_{i,k,(a,b)}$ and $t_{i,k,b,j}$. &  \hfil $2 K E \log(T_{max})$ \\
\hline
 \hfil $SRG_{i,j}$ & Indicates whether a node or edge from the $SRG_j$ is used for solution $i$. & \hfil $2$ (\# of SRGs) \\
\hline
\end{tabular}
\label{table:binaryvariables-table}
\vspace{-5mm}
\end{table}
\normalsize

In the next subsections we present two different QUBO formulations of the problem of finding multicast tree routes. 


Our two formulations present different approaches 
to generating efficient routing and spanning trees within the QUBO framework. Each approach deals with the issue of loops (see Figure \ref{fig:loops}) and time constraints (see Table \textsc{II} and Table \textsc{IV}) differently.

\small
\begin{table*}[!bp]
  \centering
\textsc{TABLE IV}\\
\textsc{Summary of constraints represented in the path-oriented approach.}
\\ \vspace{0.3cm}
\begin{tabular}{ |m{6.2cm}|m{8.7cm}|}
 \hline
\hfil Equation & \hfil Interpretation \\
\hline 
\hfil $\sum_{a\in S_k,b} e_{i,k,t_j,(a,b)} = 1$& Exactly one edge leaves each source. \\
\hline
\hfil $\sum_{a} e_{i,k,t_j,(a,t_j)} = 1$& Exactly one edge enters each sink.\\
\hline
\hfil $\forall b \notin S_k\cup T_k\quad  \sum_{a} e_{i,k,t_j,(a,b)} =\sum_{c} e_{i,k,t_j,(b,c)}$&  If one edge enters a node, one edge leaves a node.\\
\hline
\hfil $\sum_{(a,b)} T_{(a,b)}e_{i,k,t_j,(a,b)} \leq T_{max}$ & The time taken for the route from source to sink $t_k$ is less than $T_{max}$.This requires an additional $2 K \langle T_k \rangle    \log(T_{max})$ slack variables\\
\hline 
\hfil $ e_{i,k,t_j,(a,b)} \leq e_{i,k,(a,b)}$& If solution $i$ uses an edge $(a,b)$ for commodity $k$ in any route, then $e_{i,k,(a,b)}=1$.\\
\hline
\hfil $\sum_k c_k e_{i,k,(a,b)}\leq C_{(a,b)}$ & The total demand along an edge is less or equal to the capacity of that edge. This requires an additional $2 E  \langle C\rangle$ slack variables\\
\hline
$\forall  a,(a,b)\in SRG_l\quad  e_{i,k,t_j,(a,c)}\leq SRG_{i,l}$ \newline \textcolor{white}{this is an easteregg!!!}$e_{i,k,t_j,(c,a)}\leq SRG_{i,l}$ \newline\textcolor{white}{this  is an easteregg!!} $e_{i,k,t_j,(a,b)} \leq SRG_{i,l}$ & If a node or edge in $SRG_j$ is used in solution $i$, then $SRG_{i,l}=1$.\\
\hline
\hfil $ 
SRG_{0,j}+SRG_{1,j}\leq 1$ & Both solutions cannot use elements in the same risk group  $SRG_j$.\\
\hline
\hfil $e_{i,k,(a,b)}\leq e_{i,(a,b)}$ & If a commodity uses an edge $(a,b)$ for solution $i$, then $e_{i,(a,b)}=1$\\
\hline 
\hfil $e_{i,k,(a,b)}$ & If edge $(a,b)$ is used in any route for commodity $k$  to any sink for solution $i$.\\
\hline
\hfil $\sum_{(a,b)\in E} w_{(a,b)}e_{i,(a,b)} \leq W$ & The total cost of route set $i$ is below the cost threshold.\\
\hline
\end{tabular}
\label{table:constraints 2}
\end{table*}
\normalsize

\subsection{Time-Oriented Approach} \label{subsec:time_oriented}


The time-oriented approach solves directly for multicast trees. It achieves this by assigning an `time' integer variable to all nodes and constraining them such that the difference between any parent and child node to be at least the latency of that edge. We also require that all nodes which are not sources, must have a parent. Combining these, we have the value assigned to any source is an upper bound on the latency of the multicast route to any of its sinks.



Table \textsc{I} summarises the variables used to describe the problem, and Table \textsc{II} summarises the constraints for these variables. The number of variables needed for this approach with $K$ commodities and two disjoint solutions with allowed time $T_{max}$ is $\mathcal{O}(EK\log(T_{max}))$.

\begin{figure}
    \scalebox{1}{
    \includegraphics[width=\linewidth]{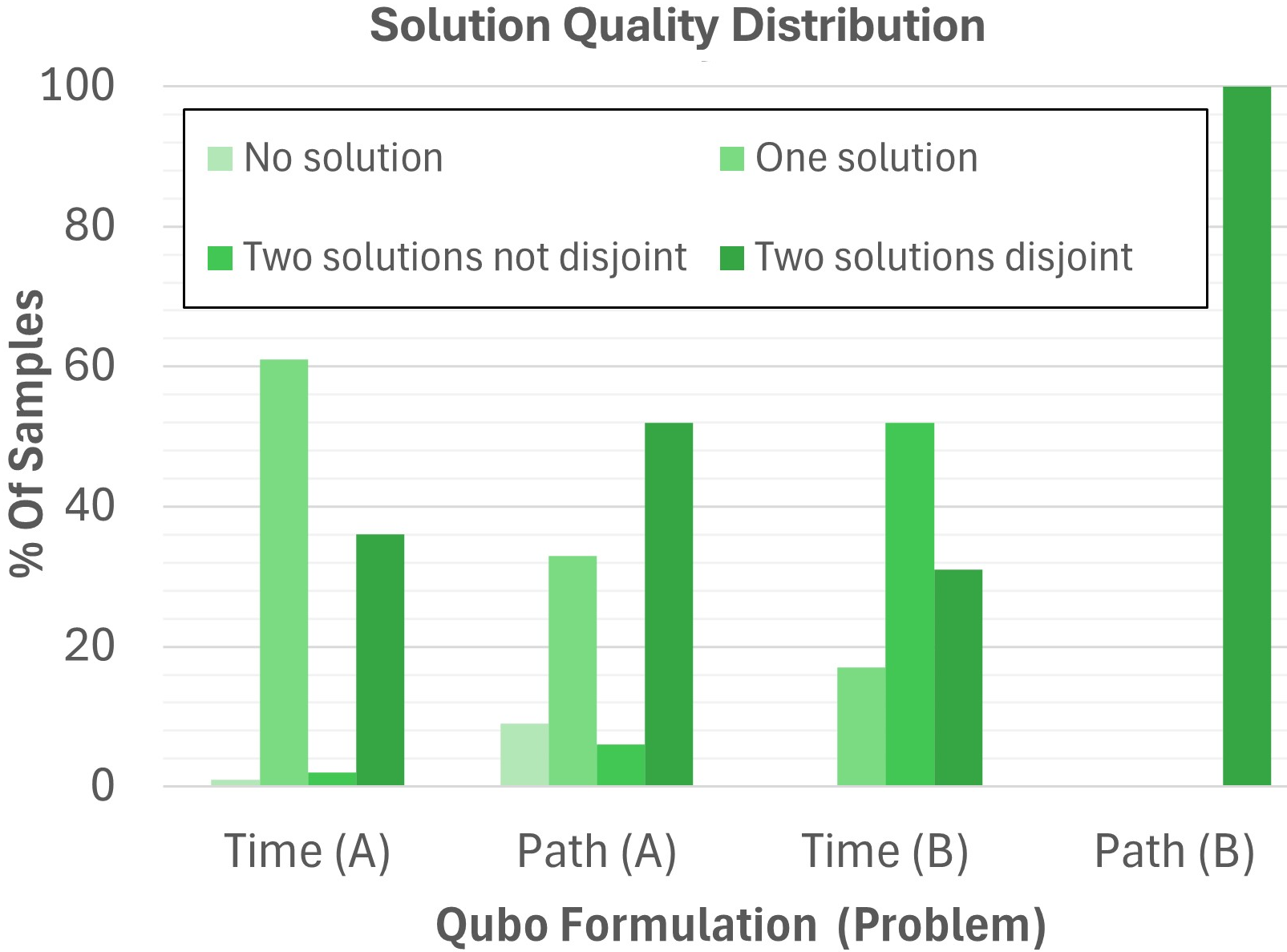}
    }
    \caption{   Solution quality for problems A and B of Figure \ref{Toy problems solution} using the Time and Path algorithms. Valid solutions are dark green.}
    \label{fig:enter-label}
    \vspace{-5mm}
\end{figure}

\subsection{Path-Oriented Approach} \label{sec:sink_oriented}

An alternative approach is to solve unicast routes and bind them into a multicast tree, using an edge cost term to improve tree efficiency. Table \textsc{III} summarises the variables used to describe the problem, and Table \textsc{IV} summarises the constraints for these variables.



This approach has the advantage over the time-oriented approach in that the `time' variables $t_{i,k,a}$ are not needed, reducing the qubit requirement for each sink. 
The number of variables needed for this approach  with $K$ commodities and two disjoint solutions with allowed time $T_{max}$ is $\mathcal{O}\left(K E\langle T\rangle+K\langle T\rangle \log{(T_{max})}\right)$. Therefore, for certain types of network (dense networks with few sinks) this approach is favourable as it requires fewer qubits than the time-oriented approach.


\small
\begin{table}
\footnotesize
\centering 
\textsc{TABLE III}\\
\textsc{Binary variables used in the path-oriented QUBO described in \cref{sec:sink_oriented}.}
\\ \vspace{0.3cm}
\begin{tabular}{|m{1.45cm}|m{3.5cm}|m{1.7cm}|}
 \hline
 \hfil Variable & \hfil Interpretation & \hfil \# of Qubits\\
 \hline
 \hfil $e_{i,k,t_k,(a,b)}$ &   Edge $(a,b)$ is used in the route for commodity $k$  to sink $t_k$ for solution $i$.  & \hfil $2 K \langle T_k \rangle E$\\
 \hline
  \hfil $e_{i,k,(a,b)}$&  Edge $(a,b)$ is used in any route for commodity $k$  to any sink for solution $i$.  & \hfil $2 K  E$\\
\hline
\hfil $e_{i,(a,b)}$&  Edge $(a,b)$ is used in solution $i$.  & \hfil $2 K  E$\\
\hline
 \hfil $SRG_{i,l}$ &  Indicates whether a node or edge from the $SRG_l$ is used for solution $i$. &   $2 (\# of SRGs)$ \\
\hline

\end{tabular}
\label{table:binaryvariables-table 2}
\vspace{-5mm}
\end{table}
\normalsize

\subsection{Routing and Wavelength Assignment (RWA)} \label{sec:wavelength_assignment}

Elastic optical networks using Wavelength Division Multiplexing (WDM) present a RWA problem \cite{simmons_routing_2020} because it is expensive and impractical to change channel wavelengths at intermediate nodes along a route. 


To solve this problem (including in the optical multicast case) we adapt the previous techniques. Here the two key constraints are that each source uses the same wavelength per commodity, and that no two commodity routing trees from different sources which share an edge  have the same wavelength (`colour').    

Tables \textsc{V} and \textsc{VI} provide details on how the `time-oriented approach' can be adapted to the colouring setting. The total number of variables (qubits) required for this formulation scales as $\mathcal{O}(CSE \log(T_{max}))$. This is polynomial qubit scaling however real instances may still require many binary variables because of the number of DWDM frequency slots and commodities (diverse information demands). There is scope for further advances to improve the scaling of this approach.

We have also extended the formulation to solve the Flexgrid Routing and Spectral Assignment (RSA) problem, as follows. For this, we define and index the set of all possible super-channels on a link. This set has cardinality $\sim$equal to the product of the total Flexgrid slots and the maximum number of slots per super-channel. We create a rule that defines whether two super-channels are blocking (i.e. based on spectral overlap). Applying this rule, we obtain the additional RSA non-blocking constraint \emph{(see last row in Table VI)} which replaces the RWA constraints in the second row, while the constraint in row 1 is replaced by a constraint that requires exactly one super-channel to correspond to commodity $k$, \emph{and} for the width of the super-channel to correspond to the bandwidth demand of commodity $k$. $c$ is now an index over all possible super-channels.


\small
\begin{table}
\footnotesize
\centering
\textsc{TABLE V}\\
\textsc{Binary variables used in the time-oriented colouring (Spectral Assignment) QUBO described in Section \ref{sec:wavelength_assignment}.}
\\ \vspace{0.3cm}
\begin{tabular}{|m{1.2cm}|m{3.4cm}|m{2.3cm}|}
 \hline
 \hfil Variable & \hfil Interpretation & \hfil \# of Qubits\\
 \hline
 \hfil $x_{s,c,a}$   &  Node $a$ is used in the tree originating from source $s$ for colour $c$.    & \hfil $CSV$\\
 \hline
 \hfil $e_{s,c,(a,b)}$&   Edge $(a,b)$ is used in the tree originating from source $s$ for colour $c$ .  & \hfil $CSE$\\
 \hline
 \hfil $t_{s,c,a,j}$ &  Time variables for each node, $$\sum_j 2^j t_{s,c,a,j} = t_{s,c,a}.$$ & \hfil $CSV \log(T_{max})$\\
 \hline
  $\xi_{s,c,(a,b),j}$ &  The product of $e_{s,c,(a,b)}$ and $t_{s,c,b,j}$. &  \hfil $CSE \log(T_{max})$ \\
\hline
 \hfil $\delta_{s,c,k}$ & Indicates whether source $s$ uses colour $c$  for commodity $k$ &  \hfil $CSK$ \\
\hline
$\zeta_{s,c,k,t}$ &  The product of $\delta_{s,c,k}$ and $x_{s,c,t}$. &  \hfil  $CK\langle ST\rangle$\\
\hline
\end{tabular}
\label{table:binaryvariables-table 3}
\vspace{-5mm}
\end{table}

\begin{table*}[!htb]
  \centering
\textsc{TABLE VI}\\
\textsc{Summary of constraints represented in the time-oriented approach for RWA and RSA problems.}
\\ \vspace{0.3cm}
\begin{tabular}{ |m{5cm}|m{9.9cm}|}
 \hline
 \hfil Equation & \hfil Interpretation\\
 \hline
\hfil  $\forall k \in S \quad \sum_{c} \delta_{s,c,k}=1$ & \emph{(only RWA)} Exactly one colour from source $s$ corresponds to commodity $k$.\\
\hline
\hfil  $\sum_{k}\delta_{s,c,k}\leq 1$& \emph{(only RWA)} Two commodities from the same source cannot use the same colour. This requires an additional $SC$ variables.\\
\hline 
\hfil $\sum_{a} e_{s,c,(a,b)} = x_{s,c,b}$& Node $b$ is connected to a tree if there is exactly one edge leading to $b$, otherwise the node is not in the tree and no edges lead to $b$.\\
\hline
\hfil $e_{s,c,(a,b)}\leq x_{s,c,a}$& A directed edge can only be used if the source node is in the tree.\\
\hline
\hfil $\begin{aligned}\xi_{s,c,(a,b),j}= e_{s,c,(a,b)}\times t_{s,c,b,j}\end{aligned}$ & $\xi_{i,k,(a,b),j}$ is the product of the two binary terms. These variables are used to construct the integer $e_{s,c,(a,b)}t_{s,c,b}= \sum_i 2^i\xi_{s,c,(a,b),j}$.\\
\hline
\hfil $t_{s,c,a} \geq e_{s,c,(a,b)}T_{(a,b)} +e_{s,c,(a,b)}t_{s,c,b} $& If edge $(a,b)$ is used, the time for information to flow from $a$ to any node $c$ beyond $b$  must be at least the time taken to flow from $b$ to $c$ plus the time taken to travel from $a$ to $b$. (This prevents loops and ensures the route is a tree). This constraint requires an additional $S C \log(T_{max})$ slack variables.\\

\hline
\hfil $\zeta_{s,c,k,t}=\delta_{s,c,k} \times x_{s,c,t}$ & $\xi_{i,k,(a,b),j}$ is the product of the two binary variables $\delta_{s,c,k}$ and $\xi_{i,k,(a,b),j}$. This is 1 exactly when the sink $t$ is part of a tree coming from source $s$ for colour $c$, and the colour $c$ from source $s$ corresponds to commodity $k$.\\
\hline
\hfil $\forall t \in T_k\quad \sum_{s,c} \zeta_{s,c,k,t}=1$ & Sink $t$ receives commodity $k$ from exactly one source through one wave length. \\
\hline
\hfil $\sum_{s}e_{s,c,(a,b)}\leq 1$& Two trees of the same colour from different sources cannot use the same edge. This requires an additional $E$ binary variables.\\
\hline
\hfil $\forall c \sum_{s, C:c\in C} e_{i,s,C,(a,b)} \leq 1$ & \emph{(only RSA)} C indexes the set of possible Superchannels, c indexes the spectral slots within C and allows rejecting solutions which are co-blocking (i.e. spectral overlap).\\
\hline

\end{tabular}
\label{table:constraints 3}
\vspace{-5mm}

\end{table*}
\normalsize

\vspace{-1mm}
\section{Experimentation}
\vspace{-2mm}
We have conducted preliminary experiments using the DWave Hybrid solver \cite{McGeoch_Pau_William_2020}. The formulations from Sections \ref{subsec:time_oriented} and \ref{sec:sink_oriented} were both successfully applied to solve the problems in Fig. \ref{Toy problems solution} (the solutions demonstrated in the figure were generated by the time-oriented approach). Since the path-oriented formalism does not necessarily yield a tree, we tested the solution resilience by requiring that if any SRG is removed from the network, at least one of the two solutions is still valid. The hyperparameters for weighting constraints were adjusted experimentally to return good solutions frequently, and the resulting distribution of solution quality is shown in Figure \ref{fig:enter-label}. In this case, the path-oriented approach was particularly successful for solving the problem of Fig 1B, yielding correct optimal solutions in 100\% of samples.




\section{Discussion}

The observed imperfect solutions may be due both to the heuristic nature of the hybrid solver, and the known physical limitations of the D-Wave 5000 QPU, including noise and fundamental scaling properties of adiabatic quantum annealing with large systems \cite{Adiabatic_Albash_2018}. It is hoped that progress in solver technology will make QUBO approaches even more practical and successful in the near future.

Different QUBO formulations resulted in both different qubit scaling, and different solution quality, for the same problem. For realistic network multicast problems, the number of sinks is typically large, which is why the time-oriented approach, which scales better with number of sinks, is preferred. However, different QUBO formulations may be more efficient for different use cases.
\IEEEpeerreviewmaketitle

\section{Conclusion}
We have shown that it is possible to formulate many optical routing problems as QUBOs to plan and optimise the transmission of multiple optical channels simultaneously on an congested network, and presented a range of use-cases (including optical unicast and multicast, RWA and RSA). We have developed two different QUBOs for planning sets of demand driven, resilient, latency-constrained, optical-multicast routes, each with different qubit-scaling functions of parameters.

Experimental testing has shown the feasibility of the `resilient multicast with constraints' formulations using the DWave Hybrid solver. Our results show that it is possible to use hybrid quantum annealers to provide good solutions. However, each run did not always generate a valid or optimal solution, so repeated sampling was required. This is likely due to the heuristic nature of the hybrid sampler, as well as noise and adiabatic energy which limits the performance of the quantum annealing component. 

The quality and optimality of solutions varied, dependent on the QUBO formulation, and the problem instance. The value of hyperparameters (weights on constraint terms) influenced the solution quality. Future work will include further benchmarking the performance, as well as experimentally evaluating the other optical routing problems (RWA and RSA) which we have formulated as QUBOs. We also plan to test all the formulations on other classes of QUBO solver.


\vspace{-1mm}
\section*{Acknowledgment}

The authors thank the UK National
Quantum Computing Centre [NQCC200921] for the hackathons that created the network to continue on the problem. BT acknowledge Innovate UK [NISQTI PN10031626].



%




\bibliography{main}{}
\bibliographystyle{unsrt}

\end{document}